\documentstyle[prb,twocolumn,aps,psfig,floats]{revtex}

\begin{document}

\title{Quantum spin chains in a magnetic field }

\author{
V. A. Kashurnikov$^1$,
N. V. Prokof'ev$^{2,3}$, B. V. Svistunov$^2$,  and M. Troyer$^4$}

\address{
$^1$ Moscow State Engineering Physics Institute,  115409, Moscow, Russia \\
$^2$ Russian Research Center ``Kurchatov Institute", 123182  Moscow, Russia \\
$^3$ Yukawa Institute for Theoretical Physics, Kyoto University, Kyoto 606-01,
Japan \\
$^4$ Institute for Solid State Physics, University of Tokyo, Roppongi
7-22-1, Minatoku, Tokyo 106, Japan}
\maketitle

\begin{abstract}
We demonstrate that the ``worm'' algorithm allows very effective and
precise quantum Monte Carlo (QMC) simulations of spin systems in a
magnetic field, and its auto-correlation time is rather insensitive to
the value of H at low temperature.  Magnetization curves for the
$s=1/2$ and $s=1$ chains are presented and compared with existing
Bethe ansatz and exact diagonalization results.  From the Green
function analysis we deduce the
magnon spectra in the $s=1$ system, and directly establish the
``relativistic" form $E(p)=(\Delta ^2 +v^2 p^2 )^{1/2} $ of the
dispersion law.
\end{abstract}


%%%%%%%%%%%%%%%%%%%%%%%%%%%%%%%%%%%%%%%%%%%%%%%%%%%%%%%%%%%%%%%%%
               \section{ Introduction }
%%%%%%%%%%%%%%%%%%%%%%%%%%%%%%%%%%%%%%%%%%%%%%%%%%%%%%%%%%%%%%%%%
\label{sec:1}

Quantum spin chains have recently been the subject of intensive
theoretical, experimental and numerical studies. Many methods were
developed and applied first to these systems.  In this paper we
present precise calculations of the magnetization curves, $m_z(H)
\equiv S_z/L$ (where $S_z=\sum_{i=1}^L s_z(i)$ is the projection of
the total spin on the direction of magnetic field, and $L$ is the
number of spins in the system), and excitation spectra in isotropic
antiferromagnetic Heisenberg chains.

The magnetization is probably the easiest quantity to measure
experimentally, and, on the contrary, the most difficult to calculate
numerically, as far as quantum Monte Carlo (QMC) methods are
concerned. In canonical ensemble (or $S_z$-conserving) schemes, $m_z$
can be obtained only from the spin and field dependence of the ground
state energies $E(S_z,H)$ with subsequent extrapolation to the
thermodynamic limit.\cite{Takahashi} Precise QMC calculations of that
kind require enormous numerical effort.\cite{Miyashita} Also, most of
the schemes rely on the Suzuki-Trotter discretization of imaginary
time,\cite{Suzuki} which introduces systematic errors. On the other
hand, the loop cluster update (LCU) algorithm is suffering from
exponential slowing down in the most interesting cases (see Sec.~II)
due to small acceptance rates in a finite magnetic
field.\cite{Evertz-review} We are not aware of any large system
simulations done using LCU in strong magnetic field.  Usually precise
calculations of the $m_z$ curves were done by exact diagonalization
(ED) \cite{Takahashi,Sakai} and density matrix renormalization group
(DMRG) methods.\cite{White}

With the development of the continuous-time ``worm" algorithm
\cite{our96,ourPLA} (it is called ``worm" algorithm because the
configuration space of the system is sampled through the worm-like
motion of world line discontinuities, or, more generally, source
operators) which works with the grand canonical ensemble (i.e.,
samples all $S_z$), the magnetization is calculated as easily as any
other ``standard" thermodynamic quantity like the energy. Furthermore,
in the same calculation one collects statistics for the Green
function, $G(r,\tau )= -{\cal T} < s^{-}(0,0)s^{+}(r,\tau )> $, where
${\cal T}$ stands for the time ordering, and $s^{\pm } = s_x \pm
is_y$.  Thus in a single QMC simulation done at low temperature one
can directly measure not just $E(T,H)$, but also $m_z(H)$, the
single-particle excitation spectrum, critical indices of $G(i,\tau )$,
the susceptibility $\chi (H) =dm_z(H)/dH$, the spin stiffness
$\Lambda_s(H)$, and more. This information may be used for precise
calculations of spin gaps and magnon dispersion curves.

In what follows we first verify, in Sec.~II, by performing an
autocorrelation times analysis, that the ``worm" algorithm does not
suffer strongly from slowing down in finite magnetic field and compare
its performance with the continuous-time LCU method for $s=1/2$
chains.  In Sec.~III we present our data for the magnetization curves
in $s=1/2$ and $s=1$ chains and compare them with the results of ED
and DMRG studies. In Sec.~IV we calculate Haldane gaps and magnon
dispersion curves for the $s=1$ system. From our data we
confirm the existence of a shallow minimum in $\chi (H)$ around
$H=3.2$ (we use units such that magnetic field, temperature, magnon
velocity, and spin gaps are measured in terms of the Heisenberg
exchange coupling constant $J$), found previously in the ED studies on
small systems.\cite{Takahashi} We exclude the possibility that this
minimum is a finite-size effect, or an artifact of extrapolation
procedure used in Ref.~\onlinecite{Takahashi}.  Our result for the
spin gap
\begin{equation}
\Delta (s=1)=0.4105(1)
\end{equation}
shows the accuracy to which this quantity may be measured in a single
QMC simulation using the ``worm" algorithm, and agrees with the ED
result $\Delta(s=1) = 0.41049(2)$\cite{Golinelli} and the DMRG results
$\Delta(s=1) = 0.41050(2)$, in Ref.~\onlinecite{White} and
$\Delta(s=1) = 0.4104892(2)$ in Ref.~\onlinecite{Qin}.  The magnon
dispersion curve fits perfectly to the relativistic form
$E^2(p)=\Delta^2+v^2p^2$, where $v$ is the magnon velocity, and $p$ is
the magnon momentum. We deduced
\begin{equation}
v(s=1)=2.48(1)
\end{equation}
from the fit; this result is as accurate as the best known DMRG values
deduced from the correlation length $v(s=1) = 2.475(5)$\cite{White-v}
and magnetization studies $v(s=1) = 2.49(1)$.\cite{Affleck-v} We also
found, that for the $s=1$ chain with $L=100$ lattice sites even
two-magnon states have visible corrections to the hard-core boson
picture which is sometimes used in fitting the data.\cite{Affleck-v}

%%%%%%%%%%%%%%%%%%%%%%%%%%%%%%%%%%%%%%%%%%%%%%%%%%%%%%%%%%%%%%%%%
\section{Critical slowing down of the ``worm'' and LCU algorithms}
%%%%%%%%%%%%%%%%%%%%%%%%%%%%%%%%%%%%%%%%%%%%%%%%%%%%%%%%%%%%%%%%%
\label{sec:2}

The LCU algorithm has been applied successfully to quantum spin systems
near phase transitions, without any sign of critical slowing
down.\cite{Troyer-cavo,Troyer-lacuo,Troyer-exp} As soon as a magnetic
field is turned on, however, the LCU updates show a severe,
exponential slowing down. The reason is that in the standard LCU
method the field cannot be taken into account when building the loop
cluster, but is treated as a global weight, modifying the flipping
probabilities of the loop clusters.

We illustrate this problem in the simplest case,
two spins in a magnetic field
\begin{equation}
H_{\rm dimer} = J{\bf S}_1{\bf S}_2 - H (S_1^z+S_2^z).
\end{equation}

\begin{figure}
\begin{center}
\mbox{\psfig{figure=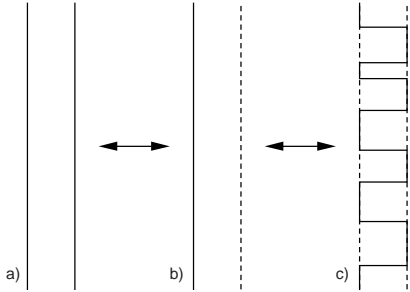,height=4cm}}
\end{center}
\caption{Illustration of the loop cluster update (LCU) algorithm
on a spin dimer.}
\label{fig:dimer-loop}
\end{figure}

\begin{figure}
\begin{center}
\mbox{\psfig{figure=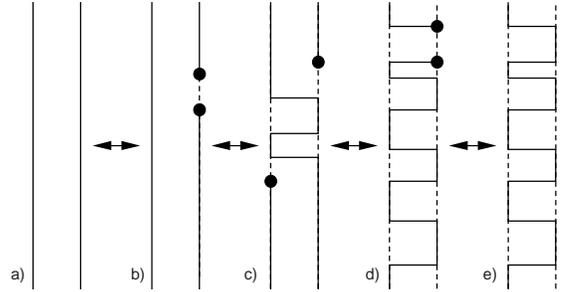,height=4cm}}
\end{center}
\caption{Illustration of the ``worm'' method
on a spin dimer.}
\label{fig:dimer-worm}
\end{figure}

At $H=J$ the ground state of this dimer system is degenerate, with the
$S^z=1$ triplet and the spin singlet both having energy $-3/4J$.  In
the world line picture the $S^z=1$ triplet is represented by two straight
world lines with $S^z_i=+1/2$ (Figs. \ref{fig:dimer-loop}a and
\ref{fig:dimer-worm}a). Starting from this configuration the LCU
method proposes to flip one of the world lines
(Fig. \ref{fig:dimer-loop}b). The new configuration however is
energetically unfavorable as its energy is $-1/4J$. It
will be accepted only with a probability $p=\exp(-\beta J/2)$, which
takes an exponential amount of time.  Once this configuration has been
accepted however it can quickly relax to the energetically favorable
singlet state, in which a world line gains exchange energy by jumping
between the two sites, as shown in
Fig. \ref{fig:dimer-loop}c. Returning from the singlet to the triplet
state is again very unlikely, since the LCU first has to create a
world line configurations with two straight world lines. Since there
is a finite probability density $(2/J)d\tau$ for having the world line
jump to the neighboring site if the two spins are antiparallel,
removing all the jumps is again exponentially rare. Thus in the LCU
for a spin-1/2 dimer at the critical point $J=H$ in a magnetic field
the autocorrelation time for the magnetization is exponentially large
\begin{equation}
\tau_{\rm dimer}^{LCU} \propto \exp{\beta H/2}.
\end{equation}
The dynamic exponent, defined as $\tau \propto \beta^z$ is
$z^{LCU}=\infty$.

\begin{figure}
\begin{center}
\mbox{\psfig{figure=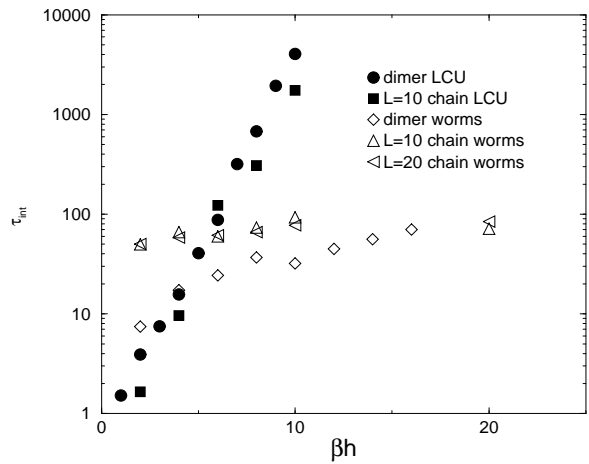,width=\hsize}}
\end{center}
\caption{Integrated autocorrelation times for the magnetization on a
$s=1/s$ dimer and a chain for both the loop cluster update
(LCU) algorithm and the
``worm'' algorithm.}
\label{fig:tau}
\end{figure}

We measured the integrated autocorrelation time using a multi cluster
loop algorithm and confirmed the exponential slowing down, as shown in
Fig. \ref{fig:tau}. In higher dimensional systems the scaling remains
exponential.

The above discussion has pointed out that in going from the triplet to
the singlet state we trade potential energy for exchange energy. The
LCU method does this in two steps, first paying a huge loss in one,
before gaining the other. The worm algorithm on the other hand can
make these trades on a local basis, as it allows configurations with
broken world lines, as is illustrated in Fig.
\ref{fig:dimer-worm}.  Starting from the triplet state the worm
algorithm flips only a short segment of the world line, with a high
acceptance rate (Fig.  \ref{fig:dimer-worm}b).  Once this short
segment has been created the world lines can gain exchange energy by
jumping between the sites, as shown in Fig.  \ref{fig:dimer-worm}c.
The worm ends make a random walk along the time direction and need a
time proportional to $\beta^2$ to wind once around the time direction
and thus change the magnetization. The autocorrelation time is
\begin{equation}
\tau_{\rm dimer}^{worm} \propto {\beta^2\over 2 \beta} \propto \beta,
\end{equation}
and the dynamical exponent $z_{\rm dimer}^{worm}=1$. The factor
$2\beta$ in the denominator normalizes the time by the minimum time
needed in any algorithm, which is proportional to the space-time volume.

We measured $\tau$ in the worm algorithm and confirmed this linear
scaling for the dimer. In the worm algorithm we measure the times in
units of $N\beta$ proposed local updates of the worm configuration,
where $N$ is the number of spins.  As shown in Fig. \ref{fig:tau},
when $\beta H>5$ the worm algorithm is much faster than the LCU
algorithm.

In higher dimensions in an $N$-spin system, the magnetization
fluctuates in general by $\sqrt{N/\beta }$.  To achieve an independent
configuration one has to again spend time \begin{equation} \tau^{worm}
\propto {(\sqrt{N/\beta }\beta)^2\over N \beta} \sim const.
\end{equation} and thus $z\approx0$.  Close to a critical point the
scaling is only slightly worse in one dimension. Modeling the magnons
as hard-core bosons and assuming a dispersion $\epsilon(k)\propto k^n$
the number of magnons at low temperatures is $M\propto N\beta^{d/n}$,
and the autocorrelation time \begin{equation} \tau_{\rm
critical}^{worm}\propto \max({(\sqrt{M}\beta)^2\over N \beta},1)
\propto \max(\beta^{1-d/n},const) \end{equation} where $d$ is the
dimension. For a one-dimensional spin chain at either critical point,
close to the spin gap $H=\Delta$ or close to the fully polarized state
$H=2zJs$ the dispersion is quadratic ($n=2$) and thus $z=1/2$. In
higher dimensions $z\approx 0$ is predicted from this
argument. Measurements of $\tau$ on a chain, also shown in
Fig. \ref{fig:tau} confirm that $\tau$ indeed increases only very
slowly with $\beta h$.

The above discussion assumed that the source operators in the worm
algorithm perform a random walk on the lattice. In the simple worm
algorithm however they do not perform a pure random walk, but are
guided by the Green function. The worm movements can be biased by the
inverse of the Green function (or an estimate thereof) though, which
makes them a random walk.

Finally we wish to note that work is in progress on improving the
scaling of the loop algorithm in a magnetic field.\cite{Evertz-Troyer}
For general couplings in the dimer case and for ferromagnetic
couplings on any lattice the exponential slowing down can be
removed. However in the most interesting case, discussed here, an
antiferromagnet on a chain or higher dimensional lattice, the slowing
down remains exponential, although the prefactor can be reduced.

%%%%%%%%%%%%%%%%%%%%%%%%%%%%%%%%%%%%%%%%%%%%%%%%%%%%%%%%%%%%%%%%%
               \section{Magnetization curves}
%%%%%%%%%%%%%%%%%%%%%%%%%%%%%%%%%%%%%%%%%%%%%%%%%%%%%%%%%%%%%%%%%
\label{sec:3}

\begin{figure}
\begin{center}
\mbox{\psfig{figure=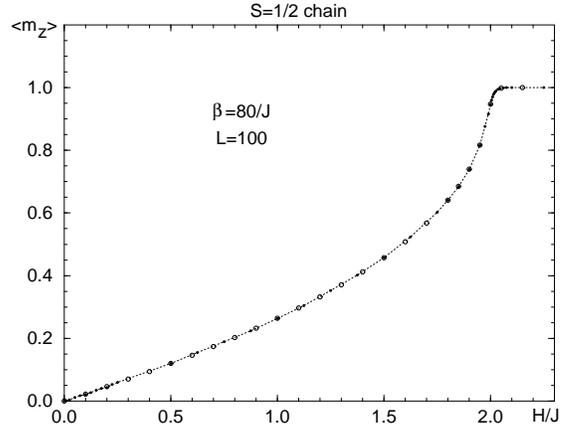,width=\hsize}}
\end{center}
\caption{
Magnetization curve for the $s=1/2$ chain.
The QMC data are shown by filled circles and the Bethe ansatz
data by open circles (the point sizes is exaggerated, especially
for the Bethe ansatz data, which were calculated up to four
meaningful digits.}
\label{fig:magnhalf}
\end{figure}

We have seen in the previous Sec. II that the ``worm" algorithm is
essentially insensitive to the value of $H$. As a testing case we
computed the magnetization curve $m_z(H)$ for the $s=1/2$ chain with
$L=100$ spins and verified that it agrees with the available Bethe
ansatz solution.\cite{s12bethe} Within our error bars the QMC and
Bethe ansatz data are in general indistinguishable when calculated at
the same value of magnetic field.  At $\beta =80$ and very small
fields $H$ one can however see in Fig. \ref{fig:magnhalf} the typical
finite-size oscillations, which appear when $\beta \sim 2\pi v /L$ and
only few magnons are present in the system; at $T=0$ these
oscillations convert into a stair-case curve.  These effects can be
reduced by going to larger $L$.  The whole curve takes about 40 hours
of CPU time on an HP-UX 9000/735.

The case of $s=1$ is more intriguing. A Bethe ansatz solution is not
available in this case, and, as mentioned above, one has to rely on ED
or DMRG. The magnetization curve for the $s=1$ chain was calculated in
Ref. \onlinecite{Takahashi}. Some aspects of this study may rise
questions. Since the largest system size was only $L=16$, the expected
square-root singularity near $H_{c1} = \Delta (s=1)$ was hardly seen,
although it is possible do recover this singularity using an
appropriate finite-size analysis.\cite{Sakai2}

Also, after extrapolating results for $E(S_z,H)$ to the thermodynamic
limit to deduce $m_z(H)$ and subsequent differentiation of $m_z(H)$, a
very shallow minimum in $\chi (H)$ around $H \approx 3.2$ was
predicted.  The amplitude of the $\chi (H)$ variation around the
minimum was only a few percent, and one might suspect that the
procedure of eliminating finite-size effects was simply not accurate
enough.

\begin{figure}
\begin{center}
\mbox{\psfig{figure=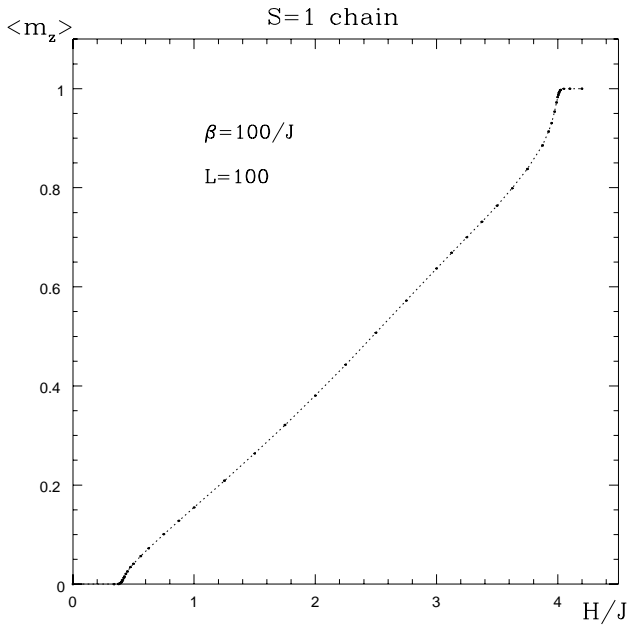,width=\hsize}}
\end{center}
\caption{Magnetization curve of the $s=1$ antiferromagnetic
Heisenberg chain.}
\label{fig:magnone}
\end{figure}

\begin{figure}
\begin{center}
\mbox{\psfig{figure=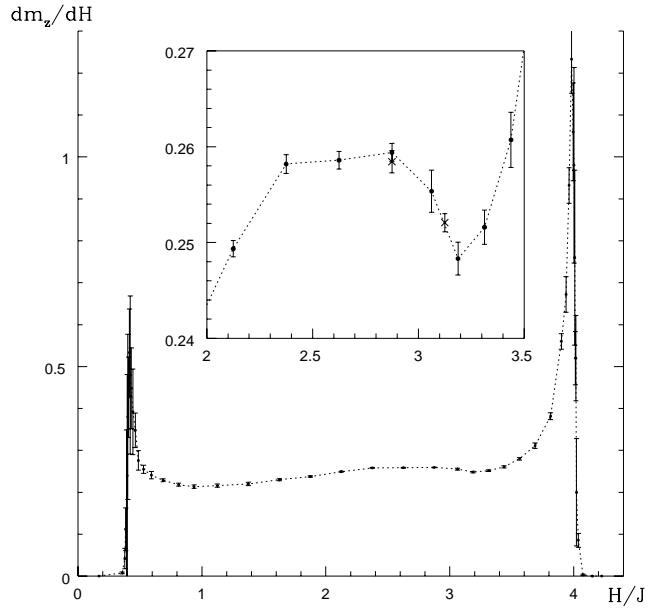,width=\hsize}}
\end{center}
\caption{Derivative of the magnetization with respect to the field
$\chi (H)$, for the $s=1$ antiferromagnetic Heisenberg chain at
$\beta=100/J$.  The region around the minimum is shown enlarged
in the inset.  Higher temperature data (at $\beta = 50/J$) are marked
by crosses.}
\label{fig:dmagnone}
\end{figure}

In Fig. \ref{fig:magnone} we show our data for $m_z(H)$ calculated
with high accuracy to ensure reliable values for the field derivative
$\chi (H)$. Obviously, there is a wide region in $H$ where $m_z(H)$ is
almost perfectly linear. In Fig. \ref{fig:dmagnone} we demonstrate
that, beyond any doubts, there is a minimum in $\chi (H)$ around $H
\approx 3.2$, and that the technique developed in
Ref. \onlinecite{Takahashi} for eliminating finite-size effects in
small clusters is remarkably accurate once the system size is larger
than the correlation length ($\xi =6.03(1)$ for the
$s=1$ chain\cite{White}). The origin of this minimum is not understood
theoretically yet.  Furthermore, we calculated $\chi (H)$ in the
region between $H=2.8$ and $H=3.2$, where $d\chi/dH$ is negative, at
higher temperature. If the minimum was any kind of finite
size effect due to level quantization, it would have been strongly
affected by temperature variation. We see, that higher-temperature
points, marked by crosses in Fig. \ref{fig:dmagnone}, are not affected
within error bars, and also give a negative $d\chi/dH$ in this region.

\begin{figure}
\begin{center}
\mbox{\psfig{figure=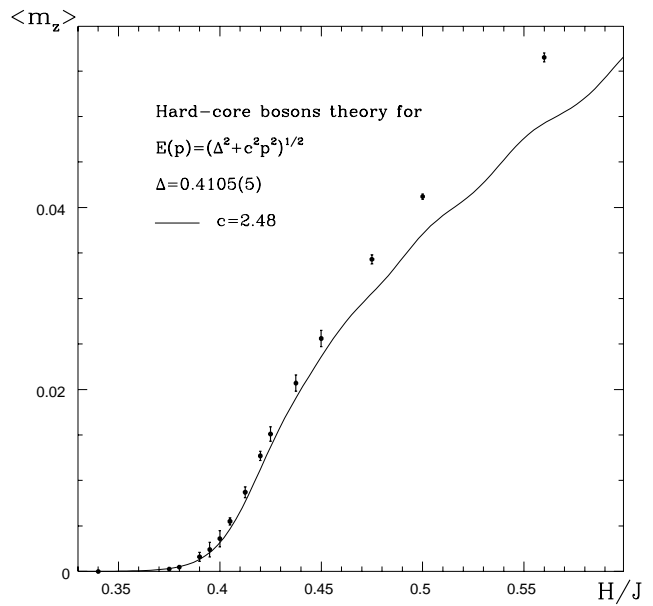,width=\hsize}}
\end{center}
\caption{Magnetization curve of the $s=1$ antiferromagnetic
Heisenberg chain close to the gap $\Delta$ on an $L=100$ site chain
at $\beta=100 $.
The solid curve shows the prediction of the hard-core boson
theory assuming $\Delta =0.4105$ and $v=2.48$.}
\label{fig:magnonefit}
\end{figure}

Clearly, for the system with $L=100$, we already see the square-root
singularity $m_z \sim \sqrt{H-\Delta }$ near the spin gap values.  It
is smeared out very close to the gap due to finite temperature. The
physics of $m_z(H)$ behavior near the gap is usually described by a
free-fermion model with periodic/antiperiodic boundary conditions
depending on the odd/even value of $S_z$ \cite{Tsvelik} or a hard-core
boson model,\cite{AffleckHC} which is exact in the dilute limit $m_z
\to 0$. It is tempting then to fit the QMC data near $H_{c1}$ to the
prediction of the hard-core boson theory at finite, but very low
temperature. In Fig. \ref{fig:magnonefit} we show the result of such a
fit, done using the following ansatz for the magnon spectrum
\begin{equation}
E(p)=\sqrt{\Delta^2+v^2p^2} \;,
\label{E(p)}
\end{equation}
with $\Delta(s=1)=0.4105$ and $v(s=1)=2.48$. Since finite size
corrections in a chain with periodic boundary conditions are
exponentially small, we used the known values of $\Delta(s=1)$ and
$v(s=1)$. Surprisingly, the fit is not good already for $m_z >0.02$,
i.e., when only two magnons are inserted into the system. Also, the
theory predicts visible finite-size oscillations in $\chi (H)$ at
$\beta =100$, which are not present in the calculated
$m_z(H)$. Although it is possible to get an almost perfect fit of the
QMC data to the hard-core boson theory up to $m \sim 0.06$ using
$v(s=1)=2.25$, we rather suggest that the difference between the
theoretical and calculated $m_z$ is due to magnon-magnon interactions
other than on-site repulsion. The gap and the magnon velocity were
deduced independently with high accuracy from the Green function (see
next Section) and exclude the possibility that $v=2.25$. It means that
one must be extremely cautious in extracting spin velocity from
multi-magnon states, i.e. states with $S_z\ge 3$, since longer range
interactions between the magnons are usually ignored in such
studies.\cite{Affleck-v}

We note, that the magnetization curve itself in Fig.
\ref{fig:magnonefit}
is sufficient to deduce the gap value rather accurately
$\Delta(s=1)=0.410(2)$ even without the Green function analysis to
which we proceed now.

%%%%%%%%%%%%%%%%%%%%%%%%%%%%%%%%%%%%%%%%%%%%%%%%%%%%%%%%%%%%%%%%%
               \section{Magnon spectra}
%%%%%%%%%%%%%%%%%%%%%%%%%%%%%%%%%%%%%%%%%%%%%%%%%%%%%%%%%%%%%%%%%
\label{sec:4}

Since the ``worm" algorithm collects ``by passing" the complete
histogram for the system's Green function, one may use it in a very
efficient way to deduce the single-particle (magnon) spectrum.  In the
present study we demonstrate how well the method works for the case
when an analytic continuation of $G(i,\tau )$ to real frequencies is
not necessary.

At very low temperature and $|H| <H_{c1}$, the statistics is
dominated by the ground-state configurations, and a piece
of an extra world line, corresponding to the $S_z=\pm 1$ states.
If $H_{c1}-H \ll H_{c1}$ and $T \ll H_{c1}-H$,
then most of the $G(p,\tau ) $ histogram
(in momentum representation) is determined by the $S_z=+1$
configurations except for very short $\sim 1$ negative times.
(since the typical length in time of an extra worldline
is of order $1/(H_{c1}-H)$, we keep $H$ rather close to the
critical field in order to make the worldline longer; this
allows one more precise determination of the Green function decay in time).
In our simulations we used $\beta =200$ and $H_{c1}-H \sim 10/\beta$.
One then expects that (counting $E(p)$ from $H$)
\begin{equation}
G(p,\tau ) \sim e^{-E(p) \tau } \;,
\label{G(p)}
\end{equation}
i.e., a simple exponential decay from which the magnon spectrum
is readily obtained as
\begin{equation}
E(p)=-{ \ln [G(p,\tau )/G(p,\tau_0 ) ] \over \tau - \tau_0 } \;.
\label{EG(p)}
\end{equation}
This definition is normalization free, although normalization
is not a problem and is fixed by the condition
$-G(i=0,\tau =+0)=s+1$.

\begin{figure}
\begin{center}
\mbox{\psfig{figure=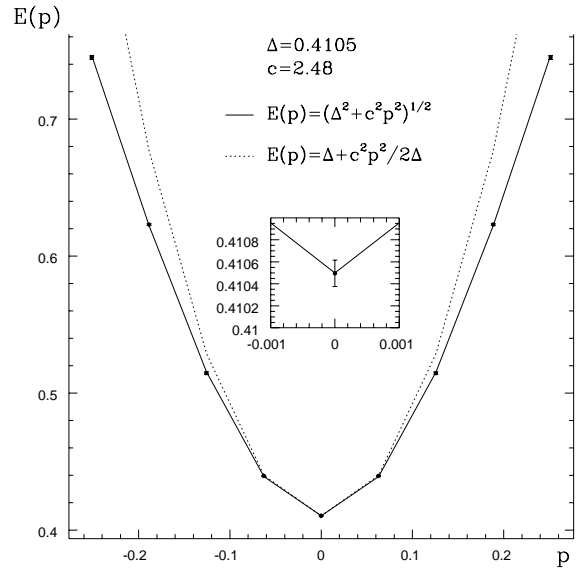,width=8cm}}
\end{center}
\caption{The magnon spectrum in the $s=1$ chain with $L=100$ spins calculated at
$\beta =200/J$. The solid line is the fit to the ``relativistic" spectrum,
and the dotted line is the parabolic approximation with the same
parameters for the gap and velocity.
The zero-momentum result is also shown in the insert.}
\label{fig:specone}
\end{figure}

First of all, we verify that the decay of $G(p,\tau )$ in time is
purely exponential for $\tau \gg 1$ until the value of $G$ becomes too
small and statistical fluctuations start dominating.  We then use
Eq.~(\ref{EG(p)}) to determine the spectrum.  For each momentum state
the reference point was set to $\tau_0 = 1/E(p)$, and then the value
of $E(p)$ was obtained from the best exponential fit to the data
between $\tau_0$ and $\tau_{max}$, where $\tau_{max}(p)$ was
close to the onset of statistical fluctuations in $G(p,\tau )$
(typically $E(p) \tau_{max} \sim 4~-~5$).
The error bars are estimated from the fluctuations of
results of up to twelve independent simulations. Our results for the
chain $s=1$ are shown in Fig. \ref{fig:specone}.
For the $s=1$ chain we studied a system with
$L=100$ spins. The precision of the data, obtained in only 50 hours of
CPU time on an HP-UX 9000/735, is sufficient to determine $\Delta
(s=1)=0.4105(1)$ --- the most accurate QMC result so far.  No finite
size corrections are necessary since these are exponentially small,
given the rather short correlation length for $s=1$. This value agrees
with the ED\cite{Golinelli} and DMRG\cite{White,Qin} data.  The magnon
velocity was found to equal $v(s=1)=2.48(1)$. This is also the most
accurate QMC result and agrees with the DMRG values $v(s=1) =2.475(5)$
in Ref.~\onlinecite{White-v}, and $v(s=1)=2.49(1)$ in
Ref.~\onlinecite{Affleck-v}.

One note is in order here. From Fig. \ref{fig:specone} we see, that
the parabolic expansion $E(p)=\Delta+v^2p^2/2\Delta$ is not accurate
for the chain $L=100$, except maybe for the first nonzero-momentum
state, but the data are consistent with the form (\ref{E(p)}) for all
the reliably determined states. This unambiguously confirms the
``relativistic'' ansatz for the dispersion law. The spin velocity is
sometimes determined from the ground-state energies $E(S_z,L)$ for
chains of comparable sizes assuming parabolic expansion and ignoring
magnon-magnon interactions (other than on-site
repulsion).\cite{Affleck-v} Since we found that both aspects have
noticeable corrections to the data in chains with $L \sim 100$, we
suspect that these corrections somehow compensate each other and the
net result turns out to come right, e.g., the DMRG result obtained in
Ref.~\onlinecite{Affleck-v} is $v=2.49(1)$. Our data give the
dispersion law directly and do not rely on any assumptions about its
form. Also, they are not affected by the magnon-magnon interactions
since the parameters $H$ and $T$ are chosen in such a way that only
single-magnon virtual states are contributing to the statistics.

%%%%%%%%%%%%%%%%%%%%%%%%%%%%%%%%%%%%%%%%%%%%%%%%%%%%%%%%%%%%%%%%%
               \section{Concluding remarks}
%%%%%%%%%%%%%%%%%%%%%%%%%%%%%%%%%%%%%%%%%%%%%%%%%%%%%%%%%%%%%%%%%
\label{sec:5}

In the present study we have shown that ``worm" algorithm is very
efficient for simulations of spin systems in finite magnetic field.
Its autocorrelation time is almost independent of $\beta $ and $H$.
For the isotropic antiferromagnetic Heisenberg chains $s=1/2$ and
$s=1$ the accuracy of our data for the magnetization curves, spin gaps
and magnon velocities is comparable with the Bethe ansatz, exact
diagonalization and DMRG results.  Moreover, from these data we found
visible longe-range (apart from onsite hard-core
repulsion) magnon-magnon interactions, and unambiguously verify the
relativistic form of the magnon dispersion law.
Our results for the magnon spectrum are the most straightforward ones,
i.e., they are not affected by the finite-size effects ($L/\xi >10$),
spin excitations at the chain ends, or magnon-magnon interactions.
Also, the numerical effort is not extreme, since one has to perform
only one simulation to get the above results. This is made possible
through the evaluation of the Green function of the system.

It was demonstrated recently how to implement the ``worm'' idea within
the framework of the LCU method.\cite{Wiese98}. Since $\Delta (s=2)$ is
rather small $\sim 0.09$ (see Refs. \onlinecite{Schollwock,Qin,Kim}),
one may deduce the gap and the spectrum very accurately
from the Green function even at $H=0$ making use of improved
estimators available in LCU algorithms \cite{Troyer}.

We would like to thank A.\ Furusaki, M.\ Sigrist, and M. Takahashi for
valuable discussions. The finite-T Bethe ansatz code for the $s=1/2$
chain was provided by M. Takahashi.  VK, NP and BS acknowledge support
from the Russian Foundation for Basic Research (VK Grant:97-02-16187,
NP and BS grant 98-02-16262), and INTAS grant 97-2124.

%%%%%%%%%%%%%%%%%%%%%%%%%%%%%%%%%%%%%%%%%%%%%%%%%%%%%%%%%%%%%


\begin{thebibliography}{99}
%%%%%%%%%%%%%%%%%%%%%%%%%%%%%%%%%%%%%%%%%%%%%%%%%%%%%%%%%%%%%
\bibitem{Takahashi}
  T. Sakai and M. Takahashi, Phys. Rev. B,
  {\bf 43}, 13383 (1991).

\bibitem{Suzuki}
   M. Suzuki, Prog. Theor. Phys. {\bf 56}, 1454 (1976);
   M. Suzuki, S. Miyashita, and A. Kuroda,
   Prog. Theor. Phys. {\bf 58}, 1377 (1977).

\bibitem{Miyashita}
      M. Roji and S. Miyashita, J. Phys. Soc. Jpn. {\bf 65}, 3317 (1996).

\bibitem{Evertz-review}
      For the latest review of the LCU methods see, e.g.,  H.G. Evertz,
      {\it The Loop Algorithm}, in {\it Numerical Methods for Lattice
      Many-Body Problems}, ed. D.J. Scalapino, Addison Wesley Longman,
      Frontiers in Physics (1998).

\bibitem{Sakai}
     T. Sakai  and M. Takahashi, cond-mat/9710327.

\bibitem{White}
      S.R. White, Phys. Rev. Lett. {\bf 69}, 2863 (1993);
      Phys. Rev. B {\bf 48}, 10345 (1993);
      S.R. White and D.A. Huse, {\it ibid.} B,
      {\bf 48}, 3844 (1993).

\bibitem{our96}
      N.V. Prokof'ev, B.V. Svistunov, and I.S. Tupitsyn, Pis'ma v Zh. Eksp.
      Teor. Fiz. {\bf 64}, 853 (1996) [JETP Lett. {\bf 64}, 911].

\bibitem{ourPLA}
      N.V. Prokof'ev, B.V. Svistunov, and I.S. Tupitsyn,
      to appear in Phys. Lett. {\bf A} (March 1998); accepted
      to Sov. Phys. JETP; cond-mat/9703200.

\bibitem{Golinelli}
      O. Golinelli, Th. Jolic{\oe}ur, and R. Lacaze, Phys. Rev. B
      {\bf 50}, 3037 (1994).

\bibitem{White-v} Spin velocity can be calculated from the
      known values of the gap and the correlation length, $v=\Delta
      \xi$. The latter was found to be equal to $\xi =6.03(1)$ in Ref.
      \onlinecite{White}.


\bibitem{Affleck-v}
      E.S. Sorensen and I. Affleck, Phys. Rev. Lett. {\bf 71},
      1633 (1993).


\bibitem{Troyer-cavo} M. Troyer, H. Kontani and K. Ueda,
     Phys. Rev. Lett. {\bf 76}, 3822 (1996).

\bibitem{Troyer-lacuo} M. Troyer, M.E. Zhitomirsky and K. Ueda, Phys.
    Rev. B {\bf 55}, R6117 (1997).

\bibitem{Troyer-exp} M. Troyer, M. Imada and K. Ueda,
    J. Phys. Soc. Jpn. {\bf 66}, 2857 (1997).

\bibitem{Evertz-Troyer} H.G. Evertz and M. Troyer, unpublished;
K. Harada and N. Kawashima, private communications.


\bibitem{s12bethe}
     M. Takahashi, Phys. Rev. B, {\bf 44}, 12382 (1991);
     the code for the finite-$T$ Bethe ansatz calculations was provided
     by M. Takahashi.

\bibitem{Sakai2}
     T. Sakai  and M. Takahashi, cond-mat/9801288.


\bibitem{Tsvelik}
      A.M. Tsvelik, Sov. Phys. JETP {\bf 66}, 221 (1987);
                    Phys. Rev. B {\bf 42}, 10499 (1990).

\bibitem{AffleckHC}
     I. Affleck, Phys. Rev. B {\bf 43}, 3215 (1991).

\bibitem{Wiese98}
      B. Brower, S. Chandrasekharan, and U.-J. Wiese,
      cond-mat/9801003.

\bibitem{Schollwock}
      U. Schollw{\"o}ck and T. Jolic{\oe}ur, Europhys. Lett.
      {\bf 30}, 493 (1995).

\bibitem{Qin}
      S. Qin, Yu-L. Liu, and  L. Yu, cond-mat/9610100.

\bibitem{Kim}
      Y.J. Kim, M. Greven, U.-J. Wiese, and R.J. Birgeneau,
      cond-mat/9712257.

\bibitem{Troyer} M. Troyer, to be published.

\end{thebibliography}
\end{document}